\documentclass[cite,figures,bm]{epl}

\title{BCS-BEC crossover in an asymmetric two-component Fermi gas}

\author{Xia-Ji Liu$^{1}$ \and Hui Hu$^{1,2}$}

\institute{$^{1}$\ ARC Centre of Excellence for Quantum-Atom
Optics, Department of Physics, University of Queensland, Brisbane,
Queensland 4072, Australia
\\
 $^{2}$\ Department of Physics, Renmin University of China, Beijing 100872, China}


\begin{document}

\maketitle

\begin{abstract}
We discuss the superfluid phase transition of a strongly interacting
Fermi gas with unequal (asymmetric) chemical potentials in two pairing
hyperfine states, and map out its phase diagram near the BCS-BEC crossover.
Our approach includes the fluctuation contributions of {}``preformed
Cooper pairs'' to the thermodynamic potential at finite temperature.
We show that, below a critical difference in chemical potentials between
species, a normal gas is unstable towards the formation of either
a finite-momentum paired Fulde-Ferrell-Larkin-Ovchinnikov superconducting
phase or a uniform superfluid, depending on the asymmetry and interaction
strengths. We determine the value of critical chemical potential mismatch,
and find that it is consistent with a recent measurement by Zwierlein
\textit{et al.} {[}Science \textbf{311}, 492 (2006){]}.
\end{abstract}

Recent experimental advances in manipulating ultracold atomic
Fermi gases via a Feshbach resonance have attracted a great amount
of interest in widely varying fields from condensed matter
physics, atomic molecular and optical physics, to particle and
astro physics \cite{crossover,ketterle,hulet}. Thanks to this
precisely controllable environment, it is now possible to
experimentally explore fundamental problems of many-body physics
in the strongly interacting region. An example of particular
interest is the understanding of fermionic pairings of
two-component Fermi gases with mismatched Fermi surfaces or
chemical potentials, in the crossover from the weak coupling
Bardeen-Cooper-Schrieffer (BCS) regime to the strongly correlated
Bose-Einstein condensate (BEC) regime
\cite{ff,liu,sarma,yip,bedaque,son,sheehy,hu,carlson}.

In standard BCS superconductivity, the presence of a Fermi surface
mismatch suppresses superfluidity, and the pairing mechanism may be
qualitatively altered. Various competing states have been proposed
within mean-field theory to describe the ground state of an asymmetric
Fermi gas in the weak coupling BCS regime: the Fulde-Ferrell-Larkin-Ovchinnikov
(FFLO) state \cite{ff}, where Cooper pairs possess a finite center-of-mass
momentum, the breached pair phase \cite{liu} and the Sarma phase \cite{sarma,yip}, 
or phase separation phases with a paired BCS superfluid being surrounded by an unpaired
normal gas \cite{bedaque}. When the interaction strength is tuned
across the BCS-BEC crossover, a rich phase diagram consisting of several
these scenarios has also been put forward \cite{son,sheehy,hu}. Very
recently, there have been two experimental investigations of fermionic
superfluidity of $^{6}$Li atoms with mismatched Fermi surfaces \cite{ketterle,hulet}.

In contrast to these earlier studies, in this paper we tackle the
crossover problem of a homogeneous asymmetric Fermi gases at finite
temperature, starting from a well-defined normal state. Our goal is
to determine the superfluid transition temperature and to obtain a
reliable phase diagram for these systems beyond mean-field. Our calculations
reveal several notable features: (\textbf{i}) The normal state is stable above a 
critical chemical potential mismatch. The critical value predicted is 
in qualitative agreement with the recent measurement. (\textbf{ii}) On 
the BCS side, the critical value is roughly proportional to the transition 
temperature of a symmetric gas and, therefore is exponentially small. 
Below it, the normal gas is unstable towards the formation of FFLO states. 
As the mismatch of Fermi surfaces decreases, a uniform BCS superfluid is 
more favorable. (\textbf{iii}) In the strong coupling BEC regime, the 
critical difference in chemical potentials is of the order of binding energy. 
The superfluid state, well described by a mixture of tightly bounded Cooper pairs
and unpaired fermions in this limit, is thereby remarkable robust. 
(\textbf{iv}) The transition temperatures at fixed chemical potential imbalances
are also determined, giving rise to a finite-temperature phase diagram.

All these results are derived below by generalizing a thermodynamic
approach by Nozi\`{e}res and Schmitt-Rink (NSR) \cite{nsr,randeria,allan}
to asymmetric Fermi gases. This approach takes into account the large
fluctuation effects that are necessary in order to capture the essential
physics at the crossover.

\textit{NSR approach to an asymmetric Fermi gas}.---Because the
Feshbach resonance of $^{6}$Li atoms used in the experiment is
extremely broad \cite{huletprl,xiaji}, we can use a single-channel
model to describe the Fermi gas across a Feshbach resonance:
\begin{equation}
{\mathcal{H}}=\sum_{{\mathbf{k}}\sigma}\left(\epsilon_{{\mathbf{k}}}-\mu_{\sigma}\right)c_{{\mathbf{k}}\sigma}^{+}c_{{\mathbf{k}\sigma}}+U\sum_{{\mathbf{k}k}^{\prime}{\textbf{q}}}c_{{\mathbf{k}}\uparrow}^{+}c_{{\textbf{q}}-{\mathbf{k}}\downarrow}^{+}c_{{\textbf{q}}-{\mathbf{k}}^{\prime}\downarrow}c_{{\mathbf{k}}^{\prime}\uparrow}.\label{hami}\end{equation}
 Here $c_{{\mathbf{k}}\sigma}^{+}$ is the creation operator for the
fermionic atoms, and the pseudospins $\sigma=\uparrow,\downarrow$
denote the two hyperfine states of $^{6}$Li. The masses are the same,
so $\epsilon_{{\mathbf{k}}}=\hbar^{2}{\mathbf{k}}^{2}/2m$ for both
species, but the chemical potentials are different, \textit{i.e.},
$\mu_{\uparrow,\downarrow}=\mu\pm\delta\mu$, to account for the asymmetry.
We will focus on the situation with fixed total particle number $n=n_{\uparrow}+n_{\downarrow}$
and fixed chemical potential difference. $U$ is the effective interaction
strength and is related to the $s$-wave scattering length $a$ via
the regularization: $(4\pi\hbar^{2}a/m)^{-1}=U^{-1}+$ $\sum_{{\mathbf{k}}}(2\epsilon_{{\mathbf{k}}})^{-1}$.
A two-body bound state arises in vacuum once the scattering length
$a$ becomes positive.

Within the NSR approach \cite{nsr}, there are two essential ingredients
in the thermodynamic potential at a temperature $T$: $\Omega(T,\mu,\delta\mu)=\Omega_{0}+\Omega_{pf}$.
Here $\Omega_{0}$ corresponds to a free Fermi gas, \begin{equation}
\Omega_{0}=\frac{1}{\beta}\sum_{{\mathbf{k}}}\ln f\left[-\left(\epsilon_{{\mathbf{k}}}-\mu_{\uparrow}\right)\right]+\ln f\left[-\left(\epsilon_{{\mathbf{k}}}-\mu_{\downarrow}\right)\right],\label{omega0}\end{equation}
 where $\beta=1/k_{B}T$ and $f\left(x\right)=[\exp(\beta x)+1]^{-1}$
is the Fermi distribution function, while $\Omega_{pf}$ is associated
with the pairing fluctuation contributions, and may be determined
by summing an infinite series of ladder diagrams \cite{nsr}, \begin{eqnarray}
\Omega_{pf} & = & \frac{1}{\beta}\sum_{{\textbf{q}},i\nu_{n}}\ln\left[-\Gamma^{-1}\left({\textbf{q}},i\nu_{n}\right)\right]e^{i\nu_{n}0^{+}},\\
 & = & -\sum_{{\textbf{q}}}\int\limits _{-\infty}^{+\infty}\frac{d\omega}{\pi}g\left(\omega\right)\delta\left({\textbf{q,}}\omega\right). \label{omegapf}\end{eqnarray}
 Here ${\textbf{q}}$ is the center-of-mass momentum, $i\nu_{n}=2\pi n/\beta$
is the bosonic Matsubara frequency,
$g\left(\omega\right)=[\exp(\beta\omega)-1]^{-1}$ is the Bose
distribution function, and the particle-particle vertex function
$\Gamma^{-1}({\textbf{q}},i\nu_{n})=1/U+\chi_{pair}^{0}$
may be written as \begin{equation}
\Gamma^{-1}=\frac{m}{4\pi\hbar^{2}a}+\sum_{{\mathbf{k}}}\left[\frac{f\left(\xi_{+}\right)+f\left(\xi_{-}\right)-1}{i\nu_{n}-2\epsilon_{{\mathbf{q/2}}}-2\epsilon_{{\mathbf{k}}}+2\mu}-\frac{1}{2\epsilon_{{\mathbf{k}}}}\right],\label{vf}\end{equation}
 with $\xi_{\pm}=\epsilon_{{\textbf{q}}/2\pm{\mathbf{k}}}-\mu\mp\delta\mu$.
Following NSR, in Eq. (\ref{omegapf}) we have converted the sum over
$i\nu_{n}$ into a contour integral, and have rewritten $\Omega_{pf}$
in terms of a phase shift defined by $\delta({\textbf{q,}}\omega)=-\mathop{\textrm{Im}}\ln[-\Gamma^{-1}({\textbf{q}},i\nu_{n}\rightarrow\omega+i0^{+})]$.
The chemical potential $\mu$ for our model is determined from the
number identity $n=-\partial\Omega/\partial\mu$, or, \begin{equation}
n=n_{F}^{0}\left(T,\mu,\delta\mu\right)+2n_{B}\left(T,\mu,\delta\mu\right),\label{number}\end{equation}
 where $n_{F}^{0}=\sum_{{\mathbf{k}\sigma}}f(\epsilon_{{\mathbf{k}}}-\mu_{\sigma})$,
and
$n_{B}=-\partial\Omega_{pf}/\partial(2\mu)=1/\pi\sum_{{\textbf{q}}}\int\nolimits
_{-\infty}^{+\infty}d\omega
g(\omega)\partial\delta({\textbf{q,}}\omega)/\partial(2\mu)$ may
be interpreted as the number of {}``preformed Cooper pairs''. The
superfluid phase transition occurs when the particle-particle
vertex function develops a pole at $i\nu_{n}=0$ for a certain
value of ${\textbf{q}}$. Therefore, the transition temperature
$T_{c}$ can be conveniently obtained by the Thouless criterion,
\begin{equation} \max\Gamma^{-1}\left({\textbf{q}},i\nu_{n}=0\right)\mid_{T=T_{c}}=0,\label{thouless}\end{equation}
 which is generalized here to take into account a nonzero center-of-mass
momentum. We must solve Eq. (\ref{number}) together with Eq.(\ref{thouless})
self-consistently, to obtain $\mu(T_{c})$ and $T_{c}$ for given
chemical potential imbalance $\delta\mu$ and interaction coupling,
where $k_{F}=(3\pi^{2}n)^{1/3}$ is the Fermi wave vector.

\textit{BCS and BEC limits}.---The NSR formalism presented above is
a simplified description of the full crossover problem for an asymmetric
Fermi gas. It is asymptotically exact in the extreme weak or strong
coupling limit ($1/k_{F}a\rightarrow\pm\infty$). In between, it is
believed to provide a qualitative interpolation scheme \cite{nsr}.

In the weak coupling limit, the phase shift $\delta({\textbf{q,}}\omega)$
is small: the number equation (\ref{number}) reduces to $n=n_{F}^{0}\left(T,\mu,\delta\mu\right)$.
The chemical potential is therefore that of a non-interacting gas
of fermions, $\mu\sim\epsilon_{F}=$ $\hbar^{2}k_{F}^{2}/2m$. Keeping
in mind that $\delta\mu,T_{c}\ll\mu$, to leading order of $1/(\beta\mu)$
and $qv_{F}{\textbf{/}}\delta\mu$, one finds the following expression
for the inverse of two-particle vertex function: \begin{equation}
\frac{2\pi^{2}\hbar^{2}}{mk_{F}}\Gamma^{-1}=\left[\ln\frac{T}{T_{c}^{0}}+h_{1}\left(\delta\mu\right)\right]-\frac{\epsilon_{F}}{6}\frac{\hbar^{2}q^{2}}{2m}h_{2}\left(\delta\mu\right),\label{vfbcs1}\end{equation}
 where $k_{B}T_{c}^{0}=k_{B}T_{BCS}=(8/\pi)e^{\gamma-2}\epsilon_{F}\exp(\pi/2k_{F}a)$
is the BCS transition temperature for a respective symmetric gas,
$h_{1}\left(x\right)=-\int_{0}^{\infty}dy\ln(\beta
y)[f^{\prime}(y+x)+f^{\prime}(y-x)]+\gamma+\ln(4/\pi)$ and
$h_{2}\left(x\right)=-\int_{0}^{\infty}dy(1/y)[f^{\prime\prime}(y+x)+f^{\prime\prime}(y-x)]$.
At small chemical potential imbalance
($\delta\mu/T_{c}^{0}\rightarrow0$),
$h_{2}\left(\delta\mu\right)>0$, and the superfluid instability
occurs at ${\textbf{q}}=0$. Asymptotically, we obtain
\begin{equation}
\frac{T_{c}\left(\delta\mu\right)}{T_{c}^{0}}=\left[1-\frac{7}{12}\left(8\zeta^{\prime}\left(-2\right)-\frac{\zeta\left(3\right)}{\pi^{2}}\right)\left(\frac{\delta\mu}{k_{B}T_{c}^{0}}\right)^{2}\right],\label{tcbcs1}\end{equation}
 where $\zeta(x)$ is the Riemann zeta function. As $\delta\mu$ increases,
$h_{2}\left(\delta\mu\right)$ crosses zero and then becomes negative.
Hence the maximum value of $\Gamma^{-1}$ locates at a nonzero center-of-mass
momentum, triggering the emergence of a non-uniform FFLO state. The
condition $h_{2}\left(\delta\mu\right)=0$ defines a tricritical point
connecting both two superfluid phases and the normal phase. We find
numerically that $\delta\mu_{tri}\simeq1.07k_{B}T_{c}^{0}\,\,$and
$T_{c,tri}\simeq0.56T_{c}^{0}$, in agreement with a previous
study from Ginzburg-Landau theory \cite{rmp}. As $\delta\mu$ increases
further and approaches a critical imbalance, the transition temperature
shrinks to zero, marking a quantum phase transition to the normal
state. In this case the expression (\ref{vfbcs1}) becomes inapplicable,
since the resulting $qv_{F}$ may become comparable with $\delta\mu$.
In the limit of $\beta\rightarrow\infty$, alternatively we may cast
Eq. (\ref{vf}) in the form, \begin{equation}
\frac{2\pi^{2}\hbar^{2}}{mk_{F}}\Gamma^{-1}=\left[\ln\frac{k_{B}T_{c}^{0}}{\delta\mu}+F\left(x\right)\right]+\frac{\pi^{2}}{6}\left(\frac{k_{B}T}{\delta\mu}\right)^{2}\frac{1}{1-x^{2}},\label{vfbcs2}\end{equation}
 where $F\left(x\right)=1-\gamma-\ln(2/\pi)-1/2\ln\left|x^{2}-1\right|-1/(2x)\ln\left|(x+1)/(x-1)\right|$
and $x=qv_{F}{\textbf{/}}\delta\mu$. The function
$F\left(x\right)$ has a maximum at $x=x_{c}\simeq1.20$, with
$F\left(x_{c}\right)\simeq0.286$. Thus the quantum phase
transition takes place at
$\delta\mu_{c}\simeq1.331k_{B}T_{c}^{0}$. Slightly below $\delta
\mu _c$ the transition temperature shows a square root behavior,
\begin{equation}
\frac{k_B T_c\left( \delta \mu \right) }{\delta \mu
_c}=\frac{\sqrt{6\left(
x_c^2-1\right) }}\pi \left( \frac{\delta \mu _c-\delta \mu }{\delta \mu _c}%
\right) ^{1/2}.  \label{tcbcs2}
\end{equation}

Deep within the BEC regime, the results are greatly affected by the
phase shift contributions. The particle-particle vertex function develops
a discrete pole at a positive frequency $\omega=\epsilon_{q}/2-2\mu+\epsilon_{b}$,
where $\epsilon_{b}=\hbar^{2}/ma^{2}$ is the binding energy. Thus,
all fermions in the less populated hyperfine state will pair up with
atoms in other state to form bound pairs. As a result, in the low
energy regime the phase shift is dominated by a bound state part,
\textit{i.e.}, $\delta(q,\omega)\simeq\pi\theta(\epsilon_{q}/2-2\mu+\epsilon_{b})$,
and the number of pairs consequently acquires a standard form of an
ideal Bose gas: $n_{B}=n_{\downarrow}\simeq\sum_{{\textbf{q}}}g(\epsilon_{{\textbf{q}}}/2-2\mu+\epsilon_{b})$.
Accordingly the chemical potential $\mu$ asymptotes to half the binding
energy. On the other hand, to accommodate the remaining free excess
fermions with density $\delta n=n_{\uparrow}-n_{\downarrow}>0$, the
chemical potential imbalance $\delta\mu$ should be necessary large.
One may anticipate that $\mu_{\downarrow}\sim-\epsilon_{b}$ while
$\mu_{\uparrow}\sim\epsilon_{F}(\delta n)$, where $\epsilon_{F}(\delta n)$
is the respective Fermi level of these excess fermions: $\delta n\approx\sum_{{\mathbf{k}}}f(\epsilon_{{\mathbf{k}}}-\mu-\delta\mu)$.
At the transition point the Thouless criterion gives rise to $\mu(T_{c})=-\epsilon_{b}/2$.
The superfluid transition temperature is therefore determined by the
strong coupling number equation describing a mixture of non-interacting
bosons and fermions, \begin{equation}
n=\sum_{{\mathbf{k}}}\frac{1}{e^{(\epsilon_{{\mathbf{k}}}-\mu_{\uparrow})/k_{B}T_{c}}+1}+2\zeta\left(\frac{3}{2}\right)\left[\frac{mk_{B}T_{c}}{\pi\hbar^{2}}\right]^{3/2}.\label{numbec}\end{equation}
 By setting $T_{c}=0$, we obtain a critical chemical potential imbalance:
$\delta\mu_{c}=\epsilon_{b}/2+2^{2/3}\epsilon_{F}$.

\begin{figure}
\onefigure [width=14cm]{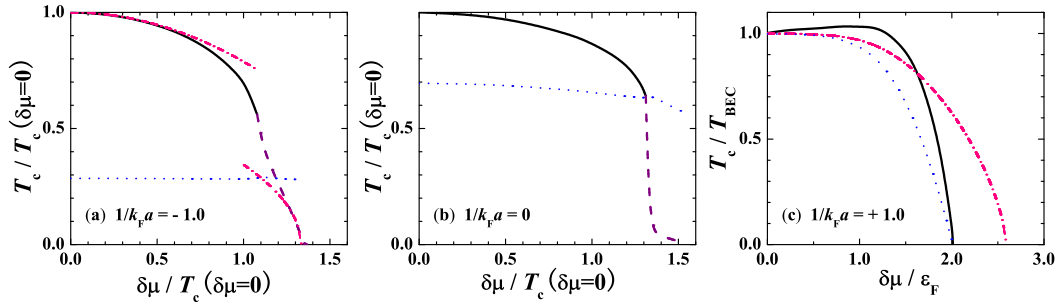} \caption{(color online)
Superfluid transition temperature $T_{c}$ as a function of the
chemical potential imbalance at various interaction strengths: (a)
$1/k_{F}a=-1$, (b) the unitary limit $1/k_{F}a=0$, and (c)
$1/k_{F}a=+1$. The phase transitions into an inhomogeneous FFLO
state have been marked by dashed lines. The dot-dashed lines in
(a) shows, respectively, the asymptotic behavior near
$\delta\mu=0$ and $\delta\mu=\delta\mu_{c}$ in Eqs. (\ref{tcbcs1})
and (\ref{tcbcs2}), while that in (c) shows the $T_{c}$ determined
from the strong coupling number equation (\ref{numbec}). The
dotted lines display the number of {}``preformed Cooper pairs''.
In (c) $T_{BEC}=(\hbar^{2}/m)\pi[n/(2\varsigma(3/2))]^{2/3}/k_{B}$
is the transition temperature for an ideal Bose gas.}
\label{fig1}
\end{figure}

Two striking features emerge from this analysis: first, in both BCS
and BEC-like phases, the superfluid state is destroyed by a sufficient
large chemical potential imbalance. On the BCS side, the critical
difference in chemical potentials is exponentially small, while deep
within the BEC regime, it is set by the binding energy. Second, in
the BCS-like phase, the preformed Cooper pairs can adjust its center-of-mass
momentum in response to the Fermi surface mismatch. This results an
inhomogeneous FFLO superfluid state in the vicinity of critical chemical
potential difference. With these background, let us turn to the numerical
NSR calculations.

\textit{Transition temperatures and phase diagrams}.---Figure 1 presents
our results for the superfluid transition temperature as a function
of the chemical potential imbalance at various interaction strengths.
The dashed lines shows the region with the onset of FFLO phases. We
plot the number of {}``preformed Cooper pairs'' in dotted lines.
Two mechanisms for the depression of transition temperatures with
increasing imbalance may be identified. On the strong coupling regime,
the decrease of $T_{c}$ is accompanied with the reduction of number
of tightly bounded pairs, as expected from the general picture of
BEC. In contrast, in the opposite BCS-like phase, the number of pairs
keeps almost constant. These pairs, resulting from the many-body effects
in this case, would be very fragile with respect to the Pauli blocking
effects arising from mismatched Fermi surfaces. We therefore explain
the reduction of $T_{c}$ as due to the loss of phase coherence between
pairs.

Figure 2a gives the critical chemical potential imbalance $\delta\mu_{c}$
throughout the BCS-BEC crossover, constituting a zero-temperature
phase diagram. For comparison, we plot the experimental data on the 
critical Fermi energy difference $\delta E_F / \epsilon _F$. These data are 
obtained indirectly by assuming a \emph{non-interacting} dispersion: 
$\delta E_F /\epsilon _F=[(1+\delta _c)^{1/3}-(1-\delta _c)^{1/3}]/2$, 
where $\delta _c$ is the measured critical population imbalance (see, \textit{i.e.}, 
the Fig. 5 in Ref. \cite{ketterle}). We find that our predictions agree 
qualitatively with the experimental results in the BCS-unitarity regime. 
However, on the BEC side they are not consistent. To understand this 
discrepancy, several remarks should be in order: First, our calculation is 
for a homogeneous gas, while the experiment is done in a trap. The presence of 
the trap tends to yield a phase separation, which may further complicate 
the comparison. Secondly, the experiment is performed at a unknown 
temperature. A close examination of the experimental data in the 
unitary limit suggests that there is an appreciable effect due to the finite 
temperature. Finally, the experimental data refer to the critical \emph{Fermi energy} 
difference, instead of the critical \emph{chemical potential} difference as 
we calculated. Only in the weakly coupling BCS regime do the chemical potentials 
equal the Fermi energies. Thus it appears to be the most serious reason responsible for the 
discrepancy between our predictions and experimental data on the BEC side. In this regard, a 
theoretical calculation for the critical population imbalance will be useful, 
enabling a direct comparison with the experiment. Unfortunately, in a region 
around the unitary limit ($-0.5<1/k_{F}a<+0.2$, see Fig. 2b), we find 
that the NSR approach generally leads to a \emph{negative} population imbalance 
at a positive chemical potential difference, implying an un-physical compressibility, \textit{i.e.},
$\partial \delta n/\partial \delta \mu <0$. This suggests the breakdown 
of the NSR treatment around the unitarity regime. We note that the negative 
compressibility is in close connection with the non-monotonous behaviour of the 
transition temperature predicted by NSR scheme for a symmetric Fermi gas \cite{nsr,randeria}. 
Presumably, the breakdown of NSR approach is due to the exclusion of the interactions between Cooper pairs, 
and may be avoided by a self-consistent improvement, 
with which the un-physical peak structure of $T_c$ is shown to be 
removed \cite{haussmann,xiaji}. In Fig. 2b, we report the critical population imbalance outside of
the unitarity regime. The two experimental points on the BEC side now become 
consistent with our theory.

\begin{figure}
\onefigure [width=13cm]{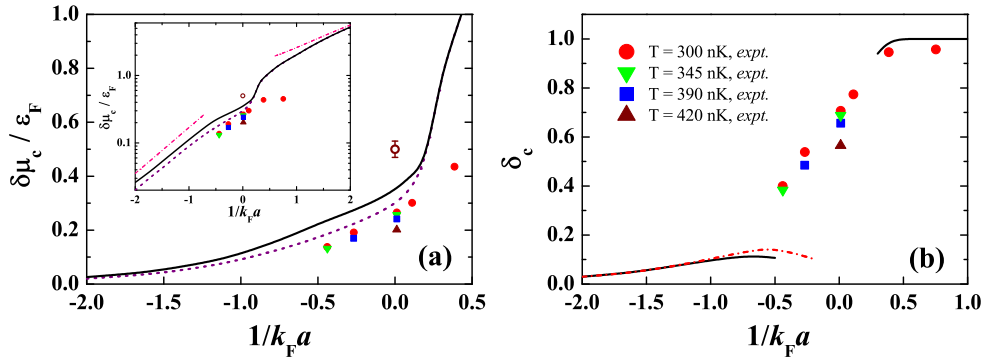} 
\caption{(color online) (a) Predicted critical chemical potential imbalance 
(solid line) in the BCS-BEC crossover. Solid symbols are the experimental 
data on the critical \emph{Fermi energy} difference. The open circle with 
an error bar in the unitary is a rough estimate from quantum Monte Carlo 
simulations in Ref. \cite{carlson}. The position of the tricritical points 
($\delta\mu_{tri}$) has been plotted by a dashed line. Inset shows 
the critical imbalance in the logarithmic scale, where two dotted-dashed 
lines represent, respectively, the critical imbalance in the extreme BCS 
and BEC limits, \textit{i.e.}, $\delta\mu_{c}\simeq1.331k_{B}T_{BCS}$ and
$\delta\mu_{c}=\epsilon_{b}/2+2^{2/3}\epsilon_{F}$. (b) Predicted critical 
population imbalance $\delta _c$ . The NSR approach (at $T=0$) breaks down around the unitarity 
regime $-0.5<1/k_{F}a<+0.2$, and thereby $\delta _c$ is unknown. The dotted-dashed line
shows the perturbation result in the BCS limit: $\delta _c=3/[2(1-2k_Fa/\pi)]\delta \mu _c$.}
\label{fig2}
\end{figure}

In Fig. 2a we have also shown the chemical potential at the 
tricritical point (with nonzero $T_{c}$) by a dashed line. Therefore, the area 
enclosed by the solid line and dashed line marks the possible region for forming
the FFLO state. The region ends up around the unitary limit. This
fact, together with the low $T_{c}$ for FFLO phases, may be used
to understand the reason of why such states are not observed in the
experiment.

\begin{figure}
\onefigure [width=7.5cm]{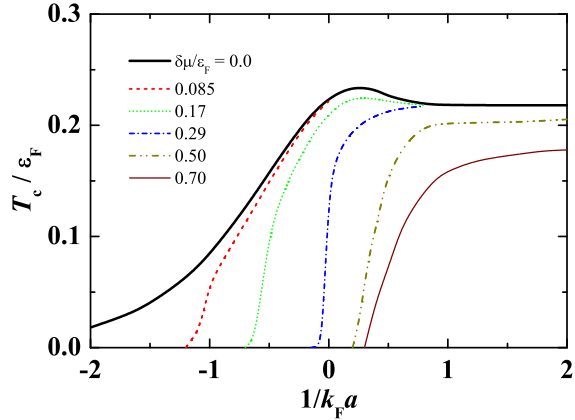} \caption{ (color online)
Finite-temperature phase diagram with the superfluid transition
temperature at the crossover plotted versus chemical potential
imbalances. For a better illustration, we have subtracted half the
binding energy from the imbalance, except for the curve with
$\delta\mu=0$ .}
\label{fig3}
\end{figure}

For completeness, in Fig. 3 we consider the transition temperature
against the interaction strength at several fixed values of chemical
potential imbalance, which makes up a finite temperature phase diagram.
In particular, there is no superfluid state on the BCS side above
$\delta\mu\simeq0.3\epsilon_{F}$.

\textit{Conclusions}.---In summary, based on a generalized NSR approximation
beyond mean-field, we have qualitatively determined the superfluid
transition temperature and a phase diagram for a two-component Fermi
gas near a Feshbach resonance with mismatched Fermi surfaces. The
resulting critical imbalance in chemical potentials agrees qualitatively well with
the recent experimental findings \cite{ketterle}. While our approach
starts from a well-defined normal state and thereby avoids the complication
of assuming a possible pairing scheme, it cannot be used to identify
the transition between different competing phases within the superfluid
domain. This fascinating issue could be addressed by extending the
current NSR approach to the broken-symmetry state \cite{hld}.

\acknowledgments
We acknowledge stimulating discussions with Professor P. D. Drummond. This
work was supported by the Australian Research Council Center of Excellence
and the National Science Foundation of China with Grant No. 10574080.

\end{document}